\documentclass[11pt,a4paper]{article}

\usepackage{jheppub}
\usepackage{graphicx}
\usepackage{float}
\usepackage{slashed}
\usepackage[font=small, labelfont=bf]{caption}
\usepackage[labelformat=simple]{subcaption}

\usepackage{amsmath, amsthm, amssymb, amsfonts}
\usepackage{multirow}
\usepackage[numbers,sort&compress]{natbib}
\usepackage{hyperref}

\newcommand{\mc}{\mathcal}
\newcommand{\be}{\begin{equation}}
\newcommand{\ee}{\end{equation}}
\newcommand{\bea}{\begin{eqnarray}}
\newcommand{\eea}{\end{eqnarray}}
\def\0{{\scriptscriptstyle 0}}
\def\32{{\scriptscriptstyle 3/2}}
\def\1{{\scriptscriptstyle 1}}
\def\ddP{{\scriptscriptstyle {\rm ddP}}}
\def\K3f{{\scriptscriptstyle {\rm K3f}}}
\def\LVS{{\scriptscriptstyle {\rm LVS}}}
\def\CY{{\scriptscriptstyle {\rm CY}}}

\def\vo{\mathcal{V}}

\allowdisplaybreaks[2]
\numberwithin{equation}{section}

\title{A geometrical upper bound on the inflaton range}

\author[a,b,c]{Michele Cicoli,}
\author[a,b]{David Ciupke,}
\author[d]{Christoph Mayrhofer,}
\author[e]{Pramod Shukla}

\affiliation[a]{\small Dipartimento di Fisica e Astronomia, Universit\`a di Bologna, \\ via Irnerio 46, 40126 Bologna, Italy}
\affiliation[b]{\small INFN, Sezione di Bologna, viale Berti Pichat 6/2, 40127 Bologna, Italy}
\affiliation[c]{\small Abdus Salam ICTP, Strada Costiera 11, Trieste 34151, Italy}
\affiliation[d]{\small Arnold Sommerfeld Center for Theoretical Physics, \\ Theresienstra{\ss}e 37, 80333 M\"unchen, Germany}
\affiliation[e]{\small Departamento de F\'{\i}sica Te\'orica and Instituto de F\'{\i}sica Te\'orica UAM/CSIC,\\
Universidad Aut\'onoma de Madrid, Cantoblanco, 28049 Madrid, Spain}

\emailAdd{mcicoli@ictp.it}
\emailAdd{ciupke@bo.infn.it}
\emailAdd{christoph.mayrhofer@physik.uni-muenchen.de}
\emailAdd{pramod.shukla@uam.es}

\abstract{We argue that in type IIB LVS string models, after including the leading order moduli stabilisation effects, the moduli space for the remaining flat directions is compact due the Calabi-Yau K\"ahler cone conditions. In cosmological applications, this gives an inflaton field range which is bounded from above, in analogy with recent results from the weak gravity and swampland conjectures. We support our claim by explicitly showing that it holds for all LVS vacua with $h^{1,1}=3$ obtained from 4-dimensional reflexive polytopes. In particular, we first search for all Calabi-Yau threefolds from the Kreuzer-Skarke list with $h^{1,1}=2$, $3$ and $4$ which allow for LVS vacua, finding several new LVS geometries which were so far unknown. We then focus on the $h^{1,1}=3$ cases and show that the K\"ahler cones of all toric hypersurface threefolds force the effective 1-dimensional LVS moduli space to be compact. We find that the moduli space size can generically be trans-Planckian only for K3 fibred examples.}


\begin{document}

\makeatletter
\let\old@fpheader\@fpheader
\renewcommand{\@fpheader}{\old@fpheader\hfill
IFT-UAM/CSIC-18-003
\hfill }

\maketitle

\bigskip

\section{Introduction}
\label{Intro}

Many recent works in relation to the weak gravity conjecture \cite{ArkaniHamed:2006dz} and the swampland conjecture \cite{Ooguri:2006in,Baume:2016psm,Klaewer:2016kiy,Blumenhagen:2017cxt} have investigated the question of both the size of moduli spaces as well as the range of validity of an effective field theory (EFT) description of trans-Planckian excursions of scalar fields in the context of the string landscape. While so far the focus has primarily been on the case of axions \cite{Rudelius:2015xta, Montero:2015ofa, Heidenreich:2015wga, Brown:2015iha, Brown:2015lia}, recently there has been interest in the axionic superpartners and, hence, in the total moduli space \cite{Palti:2017elp,Hebecker:2017lxm}. Generically, the geometric moduli space of an underlying compact geometry features at least one universal flat direction given by the modulus describing the total volume of the internal space. Without further assumptions regarding specific details of the string vacuum, this direction is limited only by EFT restrictions. Here we would like to take a different approach to this question by eliminating at least this universal flat direction via a leading order mechanism of partial moduli stabilisation which leaves at least one remaining flat direction. Thus, instead of focusing on the `full' moduli space, as it would arise in the 4-dimensional effective supergravity description of a specific string compactification, we study the `reduced' moduli space of the remaining flat directions of a string vacuum with a trustworthy EFT description. Notice that in general this `reduced' moduli space could still be non-compact with geodesic trajectories of infinite size.  

We address this question in a particular class of vacua, namely for the moduli space of K\"ahler deformations in the context of type IIB Calabi-Yau (CY) orientifold compactifications with background fluxes. This arena is well suited for this purpose since it allows for rather robust moduli stabilisation proposals. A popular mechanism for moduli stabilisation in this class of models is given by the Large Volume Scenario (LVS) which leads to phenomenologically viable vacua where the EFT approximation is expected to hold \cite{Balasubramanian:2005zx}. Moreover, for $h^{1,1}>2$, LVS models can feature a `reduced' moduli space of flat directions since leading order $\alpha'$ and non-perturbative effects can lift only the overall volume modulus and any blow-up mode resolving a point-like singularity \cite{Cicoli:2008va}. This key-property of LVS vacua has allowed very promising cosmological applications \cite{Cicoli:2011zz}. In fact, these remaining flat directions are natural inflaton candidates associated with effective shift symmetries \cite{Burgess:2014tja} which can be used to construct explicit models of large and small field inflation \cite{Conlon:2005jm, Cicoli:2008gp, Broy:2015zba, Burgess:2016owb, Cicoli:2016chb, Cicoli:2011ct, Cicoli:2015wja}. In order to trust these inflationary models, it is therefore crucial to study the size of the `reduced' LVS moduli space of inflationary flat directions.

In this paper, we shall perform this analysis by scanning for LVS vacua through the Kreuzer-Skarke list of CY threefolds with $h^{1,1}=2$, $3$ and $4$ obtained from 4-dimensional reflexive polytopes \cite{Kreuzer:2000xy}. Using a general scan based on the topology of the toric divisors, we find several LVS geometries which were so far unknown. We then focus on the $h^{1,1}=3$ cases and study their reduced moduli space $\mc{M}_r$ which turns out to be 1-dimensional. The volume of $\mc{M}_r$ can be parameterised in terms of the canonically normalised inflaton $\phi$ and depends on the moduli space metric as well as the CY K\"ahler cone. By computing different approximations to the K\"ahler cones, we show that for all $h^{1,1}=3$ cases $\mc{M}_r$ is compact and its volume is bounded from above:
\begin{equation}
\frac{\Delta \phi}{M_p} \leq c\, \ln\vo\,,
\label{Delta_bound}
\end{equation}
where $c$ is an $\mc{O}(1)$ numerical coefficient and $\vo$ is the CY volume in string units, i.e.\ ${\rm Vol}({\rm CY}) = \vo\, \ell_s^6$ with $\ell_s=2\pi\sqrt{\alpha'}$.\footnote{The fact that the volume of the reduced moduli space is logarithmically bounded should not come as a surprise since it was shown that proper distances in moduli space grow at best logarithmically \cite{Palti:2017elp}.} This result has important implications for cosmological applications of LVS vacua since it sets a geometrical upper bound on the allowed inflaton range. In particular, this upper bound sets strong constraints on inflationary models which require a trans-Planckian field range to obtain enough e-foldings of inflationary expansion. In the LVS framework, large field inflationary models with this feature involve CY geometries with a K3 fibration \cite{Cicoli:2008gp, Broy:2015zba, Burgess:2016owb, Cicoli:2016chb}. These models are particularly interesting, not just because they can predict a large tensor-to-scalar ratio of order $r\sim 0.005 - 0.01$, but also because they can be successfully embedded into global CY orientifold compactifications with explicit moduli stabilisation and chiral brane setup \cite{Cicoli:2016xae, Cicoli:2017axo}. Interestingly, we find that the size of the reduced moduli space can generically be trans-Planckian only for this kind of examples which feature a K3 fibration.

Based on our systematic analysis for $h^{1,1}=3$ and the characteristic properties of any LVS vacuum combined with the CY K\"ahler cone conditions, we expect our results to hold more generally for $h^{1,1}>3$ too. This leads us to formulate the following conjecture: \\[2mm]

\noindent\textbf{LVS moduli space conjecture:}\\
\textit{The reduced moduli space $\mc{M}_r$ of LVS vacua is compact and its volume is bounded by:
\begin{equation}
{\rm Vol}(\mc{M}_r) \lesssim \, \left[ \ln \left(\frac{M_p}{\Lambda}\right) \right]^{\rm{dim}(\mc{M}_r)}\,,
\label{Delta_bound2}
\end{equation}
where $\Lambda$ is the cut-off of the EFT.}\\[2mm]

In general the EFT cut-off $\Lambda$ is given by the Kaluza-Klein (KK) scale associated with the bulk or with some internal 4-cycle wrapped by a stack of D-branes. Expressing $\Lambda$ in terms of the CY volume (for example $\Lambda \sim M_p / \vo^{2/3}$ if we take the cut-off to be the bulk KK scale), it is immediate to realise that the more general bound (\ref{Delta_bound2}) reduces to the simpler result (\ref{Delta_bound}) valid for the 1-dimensional case. Moreover, the general bound (\ref{Delta_bound2}) is very similar to the constraints coming from the weak gravity conjecture and the swampland conjecture \cite{Baume:2016psm,Klaewer:2016kiy,Blumenhagen:2017cxt}. However, it is important to stress that, even if the bound (\ref{Delta_bound2}) is expressed in terms of the EFT cut-off, our results are more explicit since they are purely based on the geometrical features of the underlying compactification threefold. Notice however that KK and $\alpha'$ corrections modify the background geometry away from CY but we expect our main result, i.e. the compactness of the LVS reduced moduli space, to remain qualitatively correct since the exponentially large volume should help to control these corrections. This implies that the inflaton field range should be upper bounded even in the presence of a UV complete theory that would include the effect of heavy KK or stringy modes.

A simplified picture which gives a good intuition in support of our LVS moduli space conjecture is the following. In LVS vacua the leading order moduli stabilisation effects fix only the total volume modulus of the CY threefold together with the volume of a local blow-up divisor. Hence any remaining flat directions parameterising the reduced moduli space $\mc{M}_r$ correspond to divisors which can typically no longer shrink to zero or grow to infinite volume as this process would be obstructed by the presence of a blow-up divisor with fixed size inside a stabilised overall volume.

This paper is organised as follows. In Sec.~\ref{sec:setup} we explain the structure of LVS vacua which allow for a reduced moduli space whereas in Sec.~\ref{ComputeCone} we describe our computation of the K\"ahler cone of all LVS vacua with $h^{1,1}=3$ obtained from 4-dimensional reflexive polytopes. In Sec.~\ref{Results} we then show that the reduced moduli space of these string vacua is compact. This result is based both on a systematic scan through the existing list of toric CY threefolds and on an analytic proof. We finally discuss the implication of our results in Sec. \ref{Concl} and present our conclusions in Sec.~\ref{Concl2}.

\section{LVS vacua with flat directions}
\label{sec:setup}

In this section we first review the necessary ingredients for the realisation of LVS vacua which feature a reduced moduli space after leading order moduli stabilisation. We then focus on the $h^{1,1}=3$ case and describe different classes of LVS vacua depending on the geometry and topology of the underlying CY threefold.

\subsection{General conditions for LVS vacua}
\label{GenLVS}

The effective 4-dimensional supergravity obtained from type IIB CY orientifold compactifications with D3/D7-branes, O3/O7-planes and background fluxes is characterised by the following K\"ahler potential and superpotential \cite{Grimm:2004uq}:
\begin{equation}
K = - 2 \, \ln\left(\vo + \frac{\hat\xi}{2} \right) \qquad \qquad W = W_0 + \sum_{i \,\in\, I} A_i \,e^{-a_i T_i} \,,
\label{KW}
\end{equation}
where we included only the leading order $\alpha'$ correction to $K$ controlled by the parameter $\hat\xi = - \frac{\chi(X) \zeta(3)}{2 (2\pi)^3 g_s^{3/2}}$ \cite{Becker:2002nn} with $\chi(X)$ denoting the Euler number of the CY threefold $X$ and $g_s$ the string coupling. On the other hand, $W_0$ is the vacuum expectation value (VEV) of the tree-level flux superpotential \cite{Gukov:1999ya} which fixes the dilaton and the complex structure moduli \cite{Giddings:2001yu}. The superpotential includes also non-perturbative corrections which depend only on the moduli receiving instanton contributions (with index running over the subset $I$) together with model-dependent constants $A_i$ and $a_i$. The dependence of $K$ on the K\"ahler moduli is hidden inside the CY volume $\vo$. In fact, if the K\"ahler form of $X$ is expanded as $J = t^i \hat{D}_i$ in a basis of harmonic $(1,1)$-forms $\hat{D}_i$ with $i=1,...,h^{1,1}$ (we assume $h^{1,1}_+ = h^{1,1}$), the internal volume becomes:
\begin{equation}
\vo = \frac{1}{3!} \int_X J \wedge J \wedge J = \frac16 \,k_{ijk}\, t^i t^j t^k \qquad \text{where} \qquad 
k_{ijk} = \int_X \hat{D}_i \wedge \hat{D}_j \wedge \hat{D}_k \,.
\label{volForm}
\end{equation}
The corrected K\"ahler coordinates are given by $T_i = \tau_i + {\rm i}\, \int_{D_i} C_4$ and $\tau_i$ is the volume of the 4-cycle $D_i$ dual to the $(1,1)$-form $\hat{D}_i$:
\begin{equation}
\tau_i = \frac{1}{2!} \int_X \hat{D}_i \wedge J \wedge J = \frac12\, k_{ijk}\, t^j t^k\,.
\label{taui}
\end{equation}
After inverting this relation and using the volume form (\ref{volForm}), the K\"ahler potential in (\ref{KW}) can finally be expressed in terms of the K\"ahler moduli $\tau_i$. 

Focusing on the large volume regime $\vo \gg \hat\xi$ where the EFT is under control and on natural $\mc{O}(1-10)$ values of $W_0$, the generic F-term scalar potential arising from the expressions of $K$ and $W$ in (\ref{KW}) takes the form:
\begin{equation}
V = \sum_{i,j \,\in\, I} a_i a_j A_i A_j \,K^{i\bar{j}}\,\frac{e^{-\left(a_i \tau_i + a_j \tau_j\right)}}{\vo^2} 
-\sum_{i \,\in\, I} 4 A_i W_0 \,a_i\tau_i\,\frac{e^{-a_i \tau_i}}{\vo^2}
+ \frac{3\hat\xi W_0^2}{4\vo^3}\,,
\label{VF}
\end{equation}
where, without loss of generality, we consider $W_0$ and all $A_i$ to be real. Furthermore, the $C_4$-axions have been fixed already  and the inverse K\"ahler metric has the general expression \cite{Bobkov:2004cy}:
\begin{equation}
K^{i\bar{j}} = - \frac49\,  \left(\vo + \frac{\hat\xi}{2} \right) \, k_{ijk}\, t^k + \frac{4 \,\vo - \hat\xi}{\vo - \hat\xi} \, \tau_i \tau_j 
\stackrel{\vo\gg \hat\xi}{\simeq} - \frac49\,\vo \,k_{ijk}\, t^k + 4 \, \tau_i \tau_j\,.
\label{eq:Kinv}
\end{equation}
The scalar potential (\ref{VF}) admits an LVS minimum at $\vo \sim W_0 \,e^{a_i \tau_i}$ only if each of the three contributions in (\ref{VF}) has the same volume scaling. As pointed out in \cite{Cicoli:2008va}, this is guaranteed if:\footnote{As explained in \cite{Cicoli:2008va} there may exist further LVS-type vacua where the third requirement is relaxed.}
\begin{enumerate}
\item $X$ has negative Euler number $\chi(X) < 0$.\footnote{Note however that LVS vacua with $\chi(X)>0$ can exist if string loop corrections are included \cite{Cicoli:2012fh}.}

\item $X$ features at least one divisor $D_s$ which supports non-perturbative effects and can be made `small', i.e.\ the CY volume $\vo$ does not become zero or negative when $\tau_s\to 0$.

\item The element $K^{s\bar{s}}$ of the inverse K\"ahler metric scales as (for $\vo \gg \tau_s^{3/2} \sim \hat\xi$):
\begin{equation}
K^{s\bar{s}} \simeq \lambda \vo \sqrt{\tau_s} \,,
\label{Kinv}
\end{equation}
where $\lambda$ is an $\mc{O}(1)$ coefficient. 
\end{enumerate}

Combining (\ref{taui}) and (\ref{eq:Kinv}), it is easy to realise that the third condition above is equivalent to:
\begin{equation}
k_{ssi}\,k_{ssj} =  k_{sss}\, k_{sij} \qquad \forall \,i,j\,,
\label{kRel}
\end{equation}
if $k_{ssi}\neq 0$ for some $i$, otherwise we can clearly see from (\ref{eq:Kinv}) that $K^{s\bar{s}} \simeq 4\tau_s^2$. Moreover, the relation (\ref{kRel}) is equivalent also to the fact that $\tau_s$ is a perfect square since:
\begin{equation}
\tau_s = \frac12\, k_{sij}\, t^i t^j = \frac{1}{2 k_{sss}}\, k_{ssi} t^i \,k_{ssj} t^j = \frac{1}{2 k_{sss}}\left( k_{ssi} t^i \right)^2\,.
\label{square}
\end{equation}
Notice also that (\ref{kRel}) implies necessarily that $k_{sss}\neq 0$ since otherwise $k_{ssi}=0$ $\forall\,i$. The last condition in eq.~\eqref{square} is algorithmically simple to implement and will be the relevant one in the next sections. The second requirement above can be satisfied if $D_s$ is a del Pezzo (dP) divisor. Such divisors are key-ingredients for the realisation of LVS vacua \cite{Cicoli:2011it}. In fact, a dP divisor $D_s$ is rigid, i.e.\ $h^{2,0}(D_s)=0$, and without Wilson lines, i.e.\ $h^{1,0}(D_s)=0$. These are sufficient conditions to guarantee a non-vanishing $T_s$-dependent non-perturbative contribution to $W$. If in addition a del Pezzo satisfies the square relation in eq.~\eqref{square}, then it turns out that for an appropriate choice of basis this del Pezzo can 'diagonalise' the volume form in the sense that $k_{sss}\neq 0$ and $k_{ssi}=0$ for all $i \neq s$. Therefore, from now on we refer to them as 'diagonal' del Pezzos (ddP).\footnote{So far all the LVS examples found in the literature feature a diagonal dP divisor.} Moreover note that, being diagonal in the volume, $\tau_s$ is a natural `small' modulus since it can be shrunk to zero size without affecting the overall volume. This is related to the fact that diagonal dP divisors are blow-ups of point-like singularities \cite{Cicoli:2008va}. Notice however that not all dP divisors are diagonal, since for some of them it is never possible to find a basis of smooth divisors where the only non-zero intersection number is $k_{sss}$. These divisors are called `non-diagonal' dP and can be seen as resolutions of line-like singularities \cite{Cicoli:2011it}. 

To summarise we identify LVS vacua as CY geometries such that:
\begin{enumerate}
 \item $X$ is favourable and has negative Euler number $\chi(X) < 0$.
 \item $X$ features at least one del Pezzo divisor that satisfies the square relation in eq.~\eqref{square}.
\end{enumerate}
Under the above conditions, the general scalar potential (\ref{VF}) depends only on the overall volume and $n_s$ small K\"ahler moduli which control the volume of a diagonal dP divisor and support non-perturbative effects. Hence the leading order $\alpha'$-correction to $K$ and non-perturbative corrections to $W$ stabilise $n_s$ $C_4$-axions and $(n_s+1)$ 4-cycle volume moduli at \cite{Balasubramanian:2005zx}:
\begin{equation}
\frac{3\,\hat\xi}{2} \simeq \sum_{i=1}^{n_s}\sqrt{\frac{2}{d_i}}\,\tau_{\0,i}^{3/2}\qquad\text{and}\qquad
\vo_\0 \simeq \sqrt{\frac{2}{d_i}} \, \frac{W_0\sqrt{\tau_{\0,i}}}{4\, a_i A_i}\,e^{a_i \tau_{\0,i}} \quad \forall\, i=1,...,n_s\,,
\label{eq:LVS_minimum}
\end{equation}
with $d_i = (9-p_i)$ where $p_i$ is the degree of the $i$-th diagonal dP divisor (e.g.\ $d_i=9$ for a dP$_0\equiv {\mathbb P}^2$ divisor). This leaves a reduced moduli space $\mc{M}_r$ characterised by $(h^{1,1}-n_s-1)$ 4-cycle volume flat directions and $(h^{1,1}-n_s)$ axionic flat directions. Given that the axionic directions are periodic, they cannot give rise to a non-compact moduli space. We shall therefore neglect them and focus only on the remaining $(h^{1,1}-n_s-1)$ flat directions for the 4-cycle volume moduli. 

These directions can be lifted by the inclusion of additional contributions to the effective 4-dimensional scalar potential like D-terms from magnetised D-branes \cite{Dine:1987xk, Dine:1987gj}, string loop corrections \cite{Berg:2005ja, Berg:2007wt, Cicoli:2007xp, Berg:2014ama, Haack:2015pbv}, higher derivative $\alpha'$ effects \cite{Ciupke:2015msa, Grimm:2017okk}, poly-instantons \cite{Blumenhagen:2008ji, Blumenhagen:2012kz, Blumenhagen:2012ue} or non-perturbative effects for divisors rigidified by fluxes \cite{Bianchi:2011qh, Bianchi:2012pn, Louis:2012nb}. These corrections have indeed been used to achieve full closed string moduli stabilisation and to generate the inflationary potential in several LVS models. In what follows, we shall however neglect these additional effects in order to study the size of the $(h^{1,1}-n_s-1)$-dimensional reduced moduli space $\mc{M}_r$. Clearly the simplest situation is for $n_s=1$, and so $\mc{M}_r$ is non-trivial only for $h^{1,1}\geq 3$.

\subsection{Classes of LVS vacua with \texorpdfstring{$h^{1,1}=3$}{h11=3}}
\label{sec:class_LVS}

\begin{figure}[h!]
\begin{center}
\includegraphics[width=0.3\textwidth]{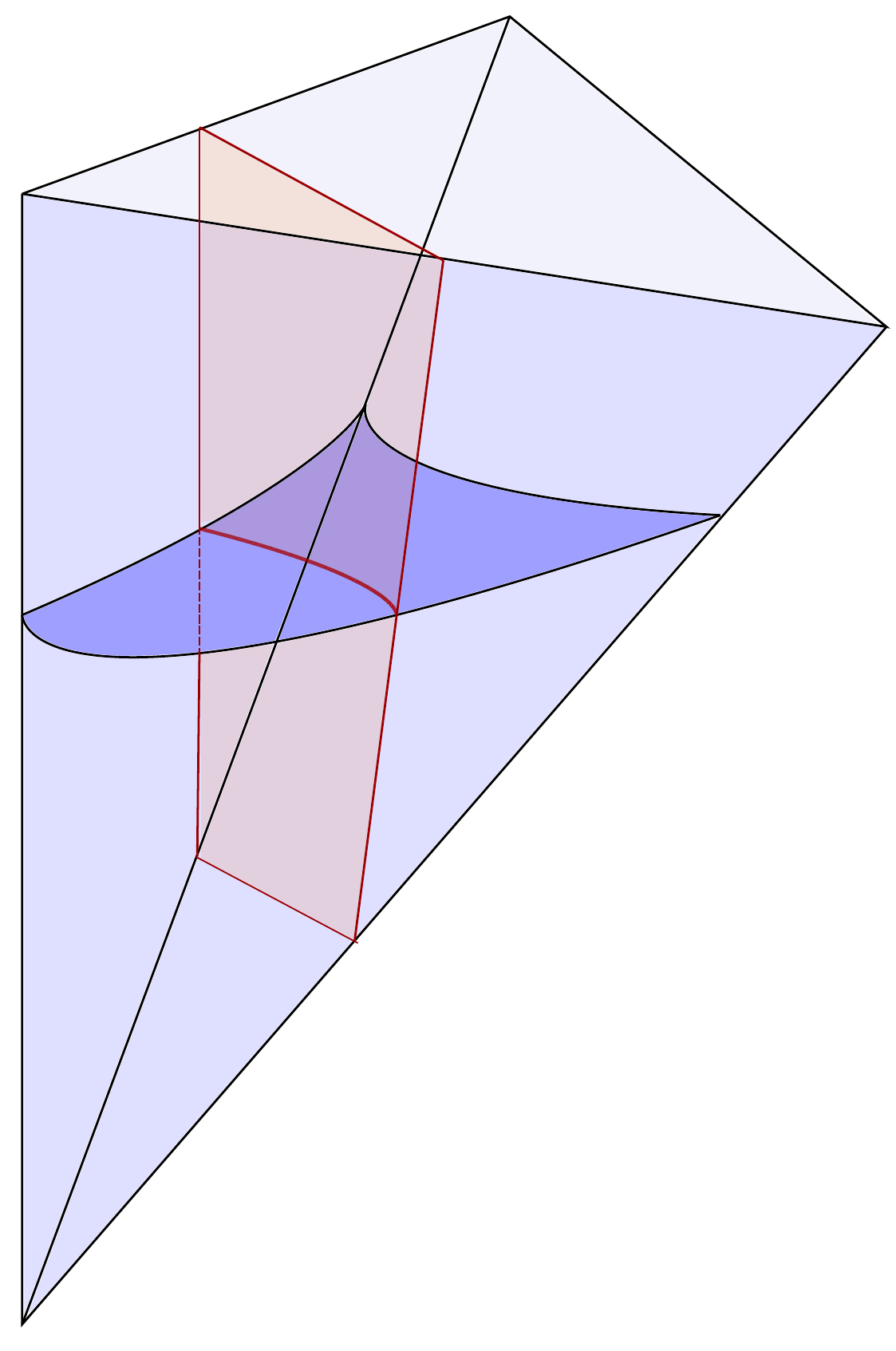}
\caption{Pictorial representation of a K\"ahler cone with $h^{1,1}=3$ parametrised by the 2-cycle volumes $t^i$. The hypersurfaces of~\eqref{eq:hypersurface} are represented respectively in blue and red. The intersection between these two hypersurfaces inside the cone represents the reduced moduli space $\mc{M}_r$.} 
\label{cone}
\end{center}
\end{figure}

The simplest LVS vacua with $h^{1,1} \geq 3$ and only $n_s=1$ feature an $(h^{1,1}-2)$-dimensional reduced moduli space $\mc{M}_r$ parameterised by the unfixed 4-cycle volume moduli. $\mc{M}_r$ can be thought of as a subspace of the full K\"ahler moduli space $\mc{M}$ defined by the hypersurface equations:
\begin{equation}
\vo(\tau_i) =  \vo_\0 \qquad\text{and}\qquad\tau_s = \tau_\0 \,,
\label{eq:hypersurface}
\end{equation} 
where $\vo_0$ and $\tau_0$ are the fixed values given in~\eqref{eq:LVS_minimum}. The geometry of the full K\"ahler moduli space $\mc{M}$ of the CY threefold $X$ is derived from the metric descending from the K\"ahler potential in (\ref{KW}) and the K\"ahler cone conditions defined as:
\begin{equation}
\int_{C_i} J > 0 \,,
\end{equation}
for all curves $C_i$ in the Mori cone of $X$, that is the cone of curves. As explained above, from now on we understand $\mc{M}$ as the $h^{1,1}$-dimensional real manifold excluding the axionic directions. A pictorial visualisation of the $\mc{M}_r$ hypersurface inside the K\"ahler cone for $h^{1,1}=3$ is given in Fig.~\ref{cone}.

Using the induced metric coming from $\mc{M}$, the volume of the $\mc{M}_r$ hypersurface can be defined as:
\begin{equation}
{\rm Vol}\,(\mc{M}_r) = \int_{\mc{M}_r} * \mathbf{1}_{r} \,.
\label{eq:M_vol}
\end{equation}
Due to the technical difficulty to compute $\mc{M}_r$, in what follows we shall perform this task only for $h^{1,1}=3$. In this case the reduced moduli space $\mc{M}_r$ turns out to be 1-dimensional. To simplify the discussion, we subdivide the LVS vacua with $h^{1,1}=3$ into different classes depending on the number $1 \leq n_\ddP \leq 2$ of diagonal dP divisors and the number $0 \leq n_\K3f \leq 1$ of K3 fibrations. Each of these classes of LVS vacua is characterised by a different volume form (\ref{volForm}) expressed in terms of the 4-cycle moduli $\tau_i$ as follows:
\begin{itemize}
\item $n_\ddP=2$ and $n_\K3f=0$: \\ 
These cases feature 2 diagonal dP moduli, and no fibred K3 divisor. The volume is completely diagonal and takes the so-called \textbf{strong Swiss cheese} form: 
\begin{equation}
\vo  = \alpha \, \tau_b^{3/2} - \beta_1 \, \tau_{s_1}^{3/2}- \beta_2 \, \tau_{s_2}^{3/2} \,,
\label{vol:2dp}
\end{equation}
where $\alpha$, $\beta_1$ and $\beta_2$ are positive and depend on the CY intersection numbers with $\beta_i$ expressed in terms of the degree of the corresponding dP divisor as $\beta_i=\frac13 \,\sqrt{\frac{2}{d_{s_i}}}$. According to the discussion in Sec.~\ref{GenLVS}, if both dP divisors support non-perturbative effects, there is no flat direction left over. We shall therefore consider only cases where at non-perturbative level $W$ depends just on a single dP modulus. This situation reproduces the case of K\"ahler moduli inflation \cite{Conlon:2005jm} which is a small field model that has been recently embedded in a chiral global construction \cite{Cicoli:2017shd} since in the inflationary regime inflaton-dependent non-perturbative effects become negligible. 

\item $n_\ddP=1$ and $n_\K3f=1$: \\ 
In these cases the CY threefold $X$ is a \textbf{K3 fibration} over a ${\mathbb P}^1$ base together with a diagonal dP divisor \cite{Cicoli:2011it}. The volume form simplifies to:
\begin{equation}
\vo = \alpha \sqrt{\tau_f} \, \tau_b - \beta \, \tau_s^{3/2} \,,
\label{vol:K3}
\end{equation}
where $\tau_f$ controls the volume of the K3 fibre and $\beta = \frac13 \,\sqrt{\frac{2}{d_s}}$ as before. This CY geometry allows the construction of several fibre inflation models depending on the different microscopic origin of the effects which lift the inflationary direction. These constructions include both small \cite{Cicoli:2011ct, Lust:2013kt} and large field models \cite{Cicoli:2008gp, Broy:2015zba, Burgess:2016owb, Cicoli:2016chb} which can be successfully embedded in global CY compactifications \cite{Cicoli:2016xae} with chiral matter \cite{Cicoli:2017axo}.

\item $n_\ddP=1$ and $n_\K3f=0$: \\ 
These cases admit only 1 diagonal dP divisor without any K3 fibration. They can be subdivided further into two subclasses depending on the structure of the volume form in the limit $\tau_s\to 0$:
\begin{enumerate}
\item For $\tau_s\to 0$, the intersection polynomial cannot be simplified by any choice of cohomology basis, and so it appears \textbf{structureless}. Hence the CY volume is:
\begin{equation}
\vo = f_\32(\tau_1,\tau_2) - \beta \, \tau_s^{3/2}\,,
\end{equation}
where $f_\32(\tau_1,\tau_2)$ is a homogeneous function of degree $3/2$ in $\tau_1$ and $\tau_2$.

\item In a proper basis of smooth divisors, when the diagonal dP divisor $\tau_s$ shrinks to zero size, the CY volume takes a simple diagonal form:
\begin{equation}
\vo = \alpha \tau_b^{3/2} - \beta_1 \, \tau_s^{3/2}- \beta_2 \left(\gamma_1\, \tau_s + \gamma_2\, \tau_*\right)^{3/2} \,,
\label{vol:ssc}
\end{equation}
with $\gamma_1$ and $\gamma_2$ positive coefficients which depend on the intersection numbers. In this case, similarly to $\tau_s$, also $\tau_b$ in (\ref{vol:ssc}) satisfies a perfect square relation. We shall call these examples as \textbf{strong Swiss cheese-like}. Notice however that the combination $\gamma_1 D_s+ \gamma_2 D_*$ does not correspond to a smooth divisor, and so there is no choice of basis of smooth divisors where $\vo$ takes the standard strong Swiss cheese (SSC) form. Moreover, even if $\vo$ looks diagonal when $\tau_s\to 0$, the divisor $D_*$ might not be shrinkable. We found by direct search that $D_*$ is rigid, i.e.\ $h^{2,0}(D_*)=0$, but it can be either a `non-diagonal' dP, a `Wilson' divisor with $h^{1,0}(D_*)\geq 1$ \cite{Blumenhagen:2012kz} or a `rigid but non-dP' divisor with Hodge diamond as a standard dP divisor but with $h^{1,1}(D_*) > 9$ \cite{Cicoli:2016xae}. Notice that cases where $D_*$ is a Wilson divisor with $h^{1,0}(D_*)= 1$ have been used to construct small field inflationary models where the inflaton potential is generated by poly-instanton effects \cite{Blumenhagen:2012ue}.
\end{enumerate}
\end{itemize}

\section{The reduced moduli space}
\label{ComputeCone}

In this section we describe how to compute the reduced moduli space for all LVS vacua with $h^{1,1}=3$. 

\subsection{Computation of the K\"ahler cone}

In the following we will be interested in CY threefolds which arise as hypersurfaces in toric varieties obtained from 4-dimensional reflexive lattice polytopes which have been classified by Kreuzer and Skarke \cite{Kreuzer:2000xy}. The geometry of the full K\"ahler moduli space $\mc{M}$, and so also of the reduced moduli space $\mc{M}_r$ and the associated volume ${\rm Vol}\,(\mc{M}_r)$, depend both on the metric as well as on the K\"ahler cone conditions. While the metric on the moduli space is readily obtained from the intersection tensor, unfortunately, there is no known algorithmic procedure to determine the exact K\"ahler cone $\mc{K}_X$ for the hypersurface CY threefolds. However, we will provide two approximate expressions for $\mc{K}$ using the following strategy: the CY Mori cone $M_X$ is contained inside a larger Mori cone $M_A$ and contains a smaller Mori cone $M_\cap$:
\begin{equation}
M_A \supseteq M_X \supseteq M_\cap \,.
\end{equation}
Conversely, the dual K\"ahler cones satisfy:
\begin{equation}
\mc{K}_A \subseteq \mc{K}_X \subseteq \mc{K}_\cap \,.
\end{equation}
If we then impose the hypersurface equations (\ref{eq:hypersurface}) which define the reduced moduli space $\mc{M}_r$, its volume turns out to have a lower and a upper bound:
\begin{equation}
{\rm Vol}\,(\mc{M}_{A,r}) \leq  {\rm Vol}\,(\mc{M}_r) \leq {\rm Vol}\,(\mc{M}_{\cap,r}) \,,
\end{equation}
where $\mc{M}_{A,r}$ and $\mc{M}_{\cap,r}$ denote the reduced moduli spaces obtained respectively from the approximated Mori cones $M_A$ and $M_\cap$. The label $A$ of the larger Mori cone $M_A$ (or the smaller K\"ahler cone $\mc{K}_A$) stays for `ambient' since $M_A$ is obtained by intersecting all the Mori cones of the ambient varieties which are connected via `irrelevant' flop transitions. These are transitions that, though changing the intersection ring of the ambient space, do not have any effect on the CY hypersurface. Therefore, we can omit curves which are flopped in these `irrelevant' transitions, when looking for the constraints on a form in the ambient variety whose pullback to the hypersurface gives a valid K\"ahler form. These irrelevant curves are discarded when we take the intersection of the Mori cones.

On the other hand, the label $\cap$ of the smaller Mori cone $M_\cap$ (or the larger K\"ahler cone $\mc{K}_\cap$) stays for `intersection' since $M_\cap$ is obtained by intersecting all toric surfaces of the ambient variety with the CY hypersurface. These toric surfaces are given by the set of the 2-dimensional cones in the fan of the toric variety. Hence, in order to obtain the curves in $M_\cap$, we have to intersect the hypersurface divisor with the two toric divisors corresponding to the two extremal rays of the 2-dimensional cones.

In the fortunate case that the larger and smaller Mori cones coincide, i.e.\ $M_A=M_\cap$, we know the actual CY K\"ahler cone by this rather simple algorithmic calculation. However, for most of the geometries this is not the case and we have to work a bit harder to obtain the actual CY K\"ahler cone (or a better approximation thereof). Given that our goal is to show that the reduced moduli space is compact, a better determination of the CY K\"ahler cone is crucial in particular for those cases where ${\rm Vol}\,(\mc{M}_{\cap,r})$ fails to be finite. The next section deals with such examples.

\subsection{Improved computation of the K\"ahler cone}
\label{sec:improv_KC}

Unfortunately, there are CY geometries for which $M_A \supsetneq M_\cap$ and ${\rm Vol}\,(\mc{M}_{\cap,r})$ is infinite and, hence, the smaller Mori cone $M_\cap$ is not sufficient to show that $\mc{M}_r$ is compact. For $h^{1,1}=3$ there are 32 toric CY hypersurfaces with $1 \leq n_\ddP \leq 2$ for which this is the case. These CY threefolds, which we have to study in more detail, fall into 3 different categories: they can be either torus fibred, K3 fibred,\footnote{In 4 cases they are both K3 and $T^2$ fibred.} or a double cover of a 3-dimensional toric variety branched over a divisor different from the anti-canonical bundle of the respective toric variety.

Since in the recent years, in the course of the F-theory phenomenology hype, all toric hypersurface $T^2$ fibrations have been worked out in detail \cite{Morrison:2012ei,Morrison:2014era,Mayrhofer:2014opa,Klevers:2014bqa}, it is rather straightforward to obtain all the curves in the fibre. The curves from the base can be pulled back via sections or multi-sections. In this way we obtain all the generating curves of the Mori cone of the CY hypersurface.

In the case of K3 fibrations, the situation is a bit more complicated because the details of all possible degenerations have not been worked out yet to the same extent as for the $T^2$ fibred examples. However, for the K3 fibrations at hand, it is not too difficult to figure out all the surfaces into which the K3 factorises at degeneration points of the fibration. With the curves from these surfaces and the pullback of the $\mathbb P^1$ of the base, we are able to obtain a much better approximation to the actual CY Mori cone than $M_\cap$.

For the CY threefolds which are double covers of toric varieties we can, first of all, employ all the effective curves of the underlying 3-dimensional toric space. To really use them for the construction of the Mori cone, we have to pull them back to the hypersurface which is a quadratic in the $\mathbb P^1$ bundle over the toric threefold, i.e.\ a double cover. In addition to these curves, there is the $\mathbb P^1$ fibre itself which is realised over the points of the base where the 3 sections which are in front of the 3 quadratic monomials vanish simultaneously.

The exact CY Mori cones for the list of these 32 CY threefolds are added as a text file to version of this paper uploaded on the \href{https://arxiv.org/}{arXiv} website.

\section{Compactness of the reduced moduli space}
\label{Results}

In this section, we present the results of our scan for toric CY threefolds with diagonal del Pezzos for $h^{1,1}=2$, $3$ and $4$. For $h^{1,1}=3$ we determine the subclasses of LVS vacua and give reference values as well as histograms for the distribution of the size of $\mc{M}_r$. In the second part, we explicitly prove that the moduli space $\mc{M}_r$ is compact for these vacua for all values of $\mathcal{V}_0, \tau_0 > 0$. 

\subsection{Scanning results}
\label{scan}

We now turn to the explicit scan for LVS vacua which, according to the discussion in Sec.~\ref{sec:setup}, we identify with CY geometries with diagonal dP divisors. Here we use the database of \cite{Altman:2014bfa} which includes CY threefolds with $h^{1,1}=2$, $3$ and $4$ built as hypersurfaces in toric varieties. For each CY threefold $X$ this database gives its topological data together with the Mori cone of the ambient space $M_A$. Since there are in general several triangulations giving rise to the same CY geometry, we do not distinguish them here and, hence, only count the number of different CY geometries $n_\CY$. We focus only on the favorable geometries as otherwise $h^{1,1}>3$. In order to compute the divisor topologies from the topological data of $X$, we use the tool \texttt{cohomCalg} \cite{Blumenhagen:2010pv, Blumenhagen:2011xn}. When searching for diagonal dP divisors, it suffices to compute the topology of toric divisors since a dP 4-cycle cannot come from a non-toric ambient space divisor. In fact, if this were the case, the dP divisor would have to be a formal sum of coordinate divisors which would allow for deformations, in clear contradiction with its rigidity. On the other hand, given that a K3 divisor can be deformed, we need to scan also for linear combinations of coordinate divisors in order to identify all possible K3 fibrations.

The general result of the scan for the total number of LVS vacua $n_\LVS$ for $h^{1,1}=2$, $3$ and $4$ is displayed in Tab.~\ref{tab:gen_res}. For the case with $h^{1,1}=3$, the number of LVS vacua for the different subclasses described in Sec.~\ref{sec:class_LVS} is given in Tab.~\ref{tab:h113}. 

\begin{table}[h!]
\centering
\begin{tabular}{|c|c|c|c||c|c|c|c|}
\hline
$h^{1,1}$ & $n_\CY$ & $n_\LVS$ & \% & $n_\ddP=1$ &  $n_\ddP=2$ &  $n_\ddP=3$  \\\hline
$2$ & $39$ & $22$ & 56.4\%	& $22$ & $-$ & $-$    \\
$3$ & $305$ & $132$ & 43.3\%	& $93$ &  $39$ & $-$  \\
$4$ & $1997$ & $749$ & 37.5\%	& $464$ &  $261$ &  $24$ \\
\hline
\end{tabular}
\caption{Number $n_\CY$ of favorable CY geometries in the database of \cite{Altman:2014bfa} for $h^{1,1}=2$, $3$, $4$ and number $n_\LVS$ of LVS vacua classified in terms of the number $n_\ddP$ of diagonal dP divisors.}
\label{tab:gen_res}
\end{table}

\begin{table}[h!]
\centering
\begin{tabular}{|c|c|c||c|c|c|c|}
\hline
$h^{1,1}$ & $n_\CY$ & $n_\LVS$ &  SSC & K3 fibred & SSC-like  & structureless   \\
\hline
$3$ & $305$ & $132$ & $39$ & $43$ & $36$ & $14$  \\
\hline
\end{tabular}
\caption{Different LVS subclasses for the cases in Tab.~\ref{tab:gen_res} with $h^{1,1}=3$.}
\label{tab:h113}
\end{table}

The next step consists in computing the reduced moduli space $\mc{M}_r$ for all LVS cases with $h^{1,1}$=3. Following the procedure described in Sec.~\ref{sec:improv_KC}, we start by computing the Mori cone $M_\cap$ given by the intersections of all toric divisors on the CY hypersurface $X$. The fixed values of $\vo_\0$ and $\tau_\0$ in \eqref{eq:LVS_minimum} determine then the two hypersurface equations \eqref{eq:hypersurface} which identify the 1-dimensional reduced moduli spaces $\mc{M}_{\cap,r}$ for $M_\cap$ and $\mc{M}_{A,r}$ for $M_A$. The volumes ${\rm Vol}\,(\mc{M}_{A,r})$ and ${\rm Vol}\,(\mc{M}_{\cap,r})$ of these two approximations of the CY reduced moduli space can finally be computed using (\ref{eq:M_vol}). 

\begin{figure}[h!]
\begin{center}
\includegraphics[width=1\textwidth]{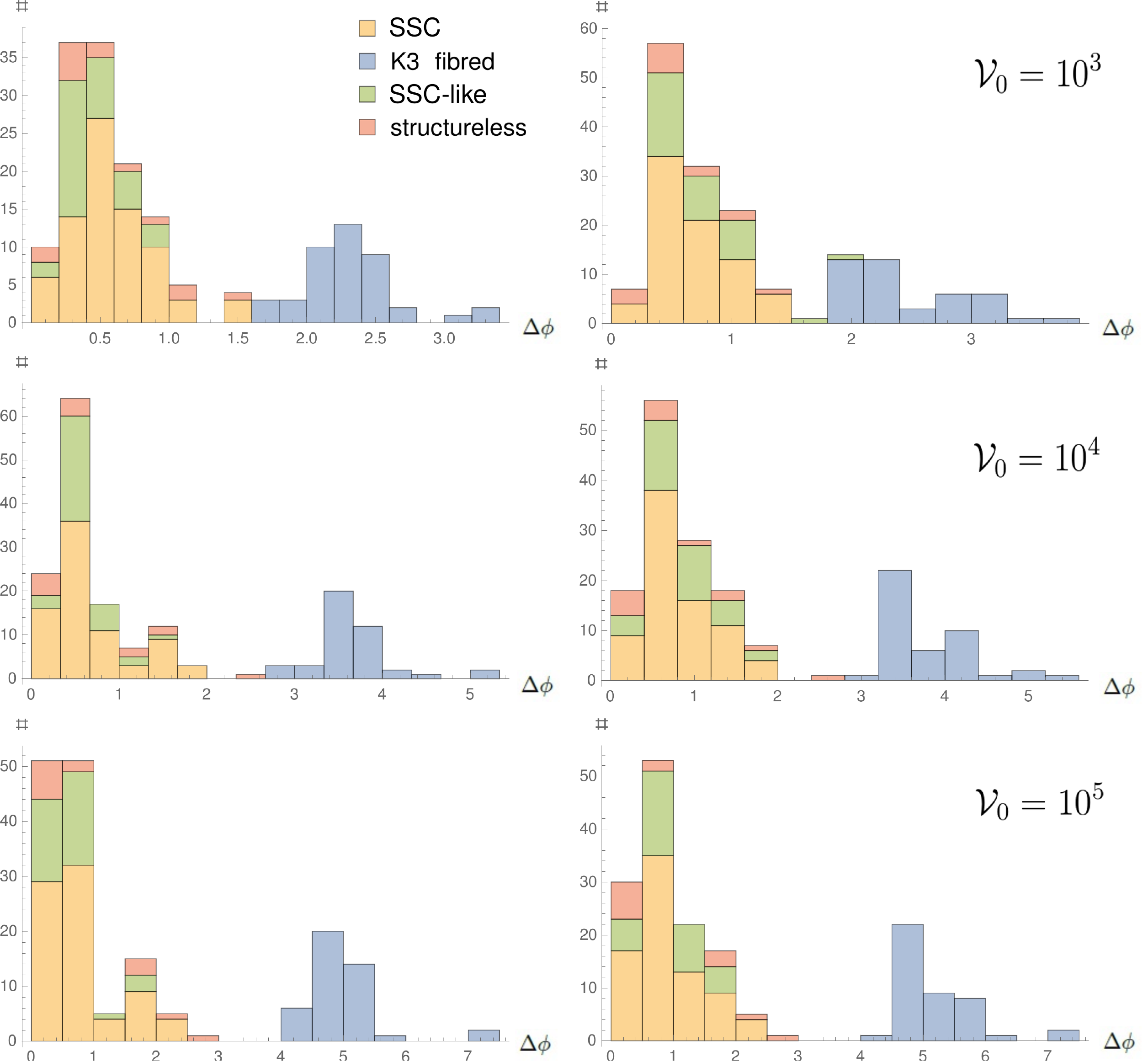}
\caption{Distribution of volumes in units of $M_p$ of reduced moduli spaces for different subclasses of LVS vacua with $h^{1,1}=3$ for $g_s=0.1$ and $\vo_\0=10^3$, $10^4$ and $10^5$. The histograms on the left display the distribution of ${\rm Vol}\,(\mc{M}_{A,r})$ whereas the ones on the right correspond to ${\rm Vol}\,(\mc{M}_{\cap,r})$.} 
\label{histolb}
\end{center}
\end{figure}

Since the size of the reduced moduli space depends on the fixed value of the overall volume (and, in a much milder way, also on $\tau_\0$), in Fig.~\ref{histolb} we show the distribution of ${\rm Vol}\,(\mc{M}_{A,r})$ and ${\rm Vol}\,(\mc{M}_{\cap,r})$ for $g_s=0.1$, which fixes $\tau_\0$ from (\ref{eq:LVS_minimum}), and different values of the CY volume $\vo_\0=10^3$, $10^4$ and $10^5$. For the 32 cases where ${\rm Vol}\,(\mc{M}_{\cap,r})$ is infinite, we replace $M_\cap$ with the exact CY Mori cone obtained by direct computation as explained in Sec.~\ref{sec:improv_KC}. For the geometries with $n_\ddP=2$ we double-count the number of vacua since, as explained in Sec.~\ref{sec:class_LVS}, the remaining flat direction can be either of the two diagonal dP blow-ups. According to Tab.~\ref{tab:h113}, this gives a total of 171 distinct LVS vacua with $h^{1,1}=3$ and 1 flat direction. 

In Tab.~\ref{tab:mean_delta} we also display the average values of ${\rm Vol}\,(\mc{M}_{A,r})$ and ${\rm Vol}\,(\mc{M}_{\cap,r})$ together with the maximal value of ${\rm Vol}\,(\mc{M}_{A,r})$ for all different LVS subclasses. Notice that we do not quote the maximal value of ${\rm Vol}\,(\mc{M}_{\cap,r})$ since it could overestimate the size of the actual CY reduced moduli space when $M_A \supsetneq M_\cap$. However, except for the $n_\ddP=2$ SSC case, for all CY geometries which have ${\rm Vol}\,(\mc{M}_{A,r})$ maximal, the Mori cone is exact since $M_A=M_\cap$. 

\begin{table}[h!]
  \centering
 \begin{tabular}{|c|c|c|c|c|c|}
    \hline
                                        & $\vo_\0$  &   SSC  & K3 fibred & SSC-like & structureless \\ \hline
                                        & $10^3$    & $0.58$ & $2.27$        & $0.43$   & $0.57$        \\
$\langle{\rm Vol}\,(\mc{M}_{A,r})\rangle$    & $10^4$    & $0.67$ & $3.62$        & $0.55$   & $0.80$        \\
                                        & $10^5$    & $0.76$ & $4.98$        & $0.62$   & $0.97$        \\ \hline
                                        & $10^3$    & $0.71$ & $2.47$        & $0.70$   & $0.57$        \\
$\langle{\rm Vol}\,(\mc{M}_{\cap,r})\rangle$ & $10^4$    & $0.81$ & $3.81$        & $0.82$   & $0.80$    \\
                                        & $10^5$    & $0.91$ & $5.17$        & $0.89$   & $0.97$    \\\hline
                                        & $10^3$    & $1.44$ & $3.31$        & $0.87$   & $1.48$     \\
${\rm max}({\rm Vol}\,(\mc{M}_{A,r}))$       & $10^4$    & $1.91$ & $5.29$        & $1.38$   & $2.41$    \\
                                        & $10^5$    & $2.38$ & $7.29$        & $1.87$   & $2.79$    \\\hline
\end{tabular}
\caption{Average volume in units of $M_p$ of the two different approximations $\mc{M}_{A,r}$ and $\mc{M}_{\cap,r}$ of the reduced moduli space of LVS vacua with $h^{1,1}=3$. We also show the maximal value of the size of the reduced moduli space $\mc{M}_{A,r}$ obtained from the ambient Mori cone $M_A$.}
\label{tab:mean_delta}
\end{table}

\begin{figure}[htb!]
\begin{center}
\includegraphics[width=1\textwidth]{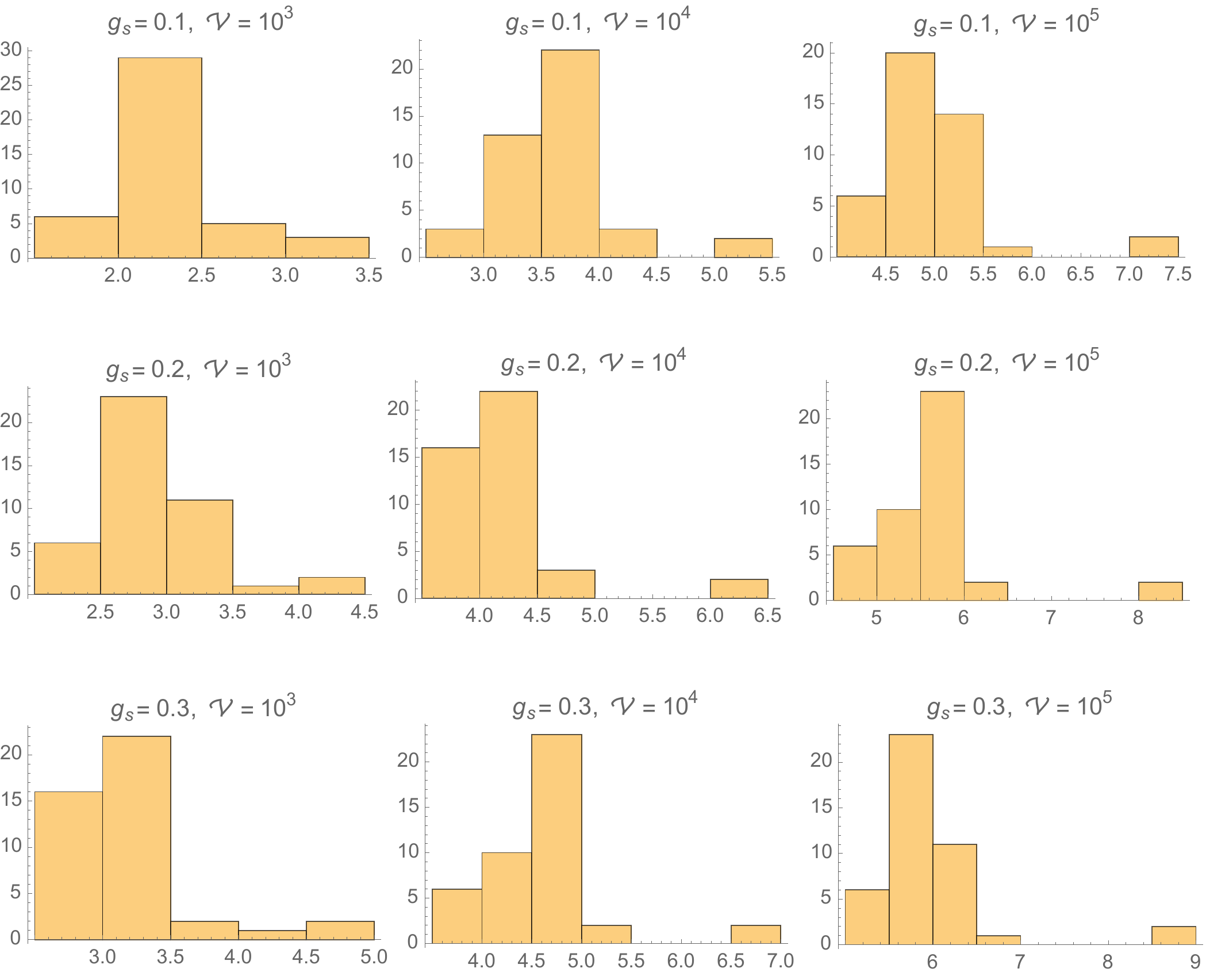}
\caption{Distribution of volumes in units of $M_p$ of the reduced moduli space $\mc{M}_{A,r}$ for all 43 K3 fibred LVS vacua for $g_s=0.1$, $0.2$ and $0.3$, and $\vo_\0=10^3$, $10^4$ and $10^5$.} 
\label{histogs}
\end{center}
\end{figure}

Tab.~\ref{tab:mean_delta} shows also that the size of the reduced moduli space varies significantly with $\vo_\0$ only for K3 fibred geometries. In this cases it is therefore important to include also the dependence of the field range $\Delta\phi$ on the fixed value of the blow-up mode $\tau_\0$. This effect can be taken into account simply considering different values of $g_s$ since, as can be seen from (\ref{eq:LVS_minimum}), $\tau_\0$ is determined just by the string coupling and the underlying CY topology. Fig.~\ref{histogs} shows the distribution of ${\rm Vol}\,(\mc{M}_{A,r})$ for all 43 K3 fibred LVS geometries with $h^{1,1}=3$ for $g_s=0.1$, $0.2$ and $0.3$, and $\vo_\0=10^3$, $10^4$ and $10^5$. The field range $\Delta\phi$ takes the largest maximal value for $g_s=0.3$ and $\vo_\0=10^5$, signaling the presence of a lower bound proportional to $\tau_\0$ (which decreases when $g_s$ increases) and an upper bound proportional to $\ln\vo_\0$. We shall confirm this result with an analytical calculation in the next section.

\subsection{Analytic proof}
\label{sec:proof_comp}

Let us now provide an analytical proof of the compactness of the reduced moduli space $\mc{M}_r$ for all subclasses of LVS vacua with $h^{1,1}=3$. Since $\mc{M}_r$ depends through the hypersurface equation on the values $\mathcal{V}_0, \tau_0$ it is important to demonstrate this for all values $\mathcal{V}_0 , \tau_0 > 0$. For SSC, K3 fibred and SSC-like LVS geometries, ${\rm Vol}(\mc{M}_r)$ can be shown to be finite by directly computing the metric on the reduced moduli space and the associated field range $\Delta\phi$. We achieve this result starting from the line element of the full moduli space obtained from the K\"ahler potential in (\ref{KW}) together with the relevant volume form as follows:
\begin{equation}
{\rm d}s^2 = g_{ij} \,{\rm d}\tau_i {\rm d}\tau_j\qquad\text{with}\qquad g_{ij} = 2 \,\frac{\partial^2 K}{\partial T_i\partial \bar{T}_j}=\frac12 \frac{\partial^2 K}{\partial \tau_i\partial \tau_j}\,,
\label{ds}
\end{equation}
where the factor $2$ in front of the metric is needed to match the standard definition of the kinetic Lagrangian $\mc{L}_{\rm kin}= \frac{\partial^2 K}{\partial T_i\partial \bar{T}_j}\,\partial_\mu T_i\partial^\mu \bar{T}_j=\frac12 \,g_{ij}\partial_\mu \tau_i \partial^\mu \tau_j$ (neglecting the axionic terms). Given that the overall volume and the diagonal dP modulus are fixed respectively at $\vo=\vo_\0$ and $\tau_s=\tau_\0$, it is useful to trade ${\rm d}\tau_b$ for ${\rm d}\vo$ via ${\rm d}\vo = \frac{\partial \vo}{\partial \tau_i}\,{\rm d}\tau_i$ and finally set ${\rm d}\vo= {\rm d}\tau_s=0$. Then (\ref{ds}) reduces to the line element ${\rm d}s_r^2$ of the 1-dimensional reduced moduli space parametrised by the remaining flat direction $\tau$ which, depending on the particular LVS subclass, can be either another diagonal dP blow-up, a K3 fibre or a modulus with a more complicated topology. The field range in units of $M_p$ can then be computed as:
\begin{equation}\label{field-range}
\Delta\phi = \int_{\tau_{\rm min}}^{\tau_{\rm max}} {\rm d}s_r\,,
\end{equation}
where the minimum and maximal values of $\tau$ are derived from the K\"ahler cone conditions of either $M_A$ or $M_\cap$ as discussed in Sec.~\ref{scan}. Note that in general we need to choose several charts and corresponding transition functions to rewrite eq.~\eqref{eq:M_vol} in terms of local coordinates on $\mc{M}_r$ and, thus, eq.~\eqref{field-range} holds only if the moduli space can be covered by a single chart. For the vacua we are considering this approach is justified for all SSC, K3 fibred and SSC-like cases, but not for the structureless geometries. For all SSC, K3 fibred and SSC-like cases we also find that:
\begin{equation}
\tau_{\rm min} = a \,\tau_\0 \qquad\text{and}\qquad \tau_{\rm max} = f_\1(\vo_\0,\tau_\0) \,,
\label{eq:field_range}
\end{equation}
where $a\geq 0$ and $f_\1$ is a homogeneous function of degree 1 in the 4-cycle moduli that is strictly positive for all $\vo_\0>0$ and $\tau_\0 >0$. 

Let us now present the metric on $\mc{M}_r$ and the associated field range for these 3 subclasses of LVS vacua with $h^{1,1}=3$: 
\newpage
\begin{itemize}
\item \textbf{SSC vacua} with $n_\ddP=2$ and $n_\K3f=0$:

Using the volume form in eq.~\eqref{vol:2dp}, the line element of the reduced moduli space parametrised by $\tau = \tau_{s_2}$ is given by:
\begin{equation}
{\rm d}s_r^2 = \frac{3 \beta_2}{4 \vo_\0  \sqrt{\tau}}\left(\frac{1 + \beta_1 \, \epsilon}{1 + \beta_1\,\epsilon + \beta_2\,  \frac{\tau^{3/2}}{\vo_\0}}\right) \,{\rm d} \tau^2\,\qquad\text{with}\qquad \epsilon \equiv \frac{\tau_\0^{3/2}}{\vo_\0} \ll 1\,.
\label{dsSSC}
\end{equation} 
In this case $a=0$ for each toric CY we have analysed, and so at one of the boundaries of $\mc{M}_r$ we find a singularity. In the full moduli space this corresponds to the face of the cone where the respective dP divisor shrinks to zero size. Despite that, the volume of the reduced moduli space becomes:

\begin{equation}
\Delta \phi = \frac{2}{\sqrt{3}} \sqrt{1 +\beta_1 \,\epsilon } \,{\rm arcsinh} \left( f_\1^{3/4} \sqrt{\frac{\beta_2}{\vo_\0 \left(1+ \beta_1 \,\epsilon\right)}} \right),
\label{DphiSSC}
\end{equation}
which, in turn, implies that $\Delta \phi$ is finite for all $\mathcal{V}_0,\tau_0>0$. Hence, in these cases $\mc{M}_r$ is compact due to the $1/\sqrt{\tau}$ behaviour of the metric near the singularity. Notice also that for $f_\1 \rightarrow \infty$ the volume of $\mc{M}_r$ behaves logarithmically and, thus, satisfies the general bound~\eqref{Delta_bound}.

\item \textbf{K3 fibred vacua} with $n_\ddP=1$ and $n_\K3f=1$:

From the volume form (\ref{vol:K3}) and parameterising the flat direction with the K3 modulus, i.e.\ $\tau = \tau_f$, we find that:
\begin{equation}
{\rm d}s_r^2 = \frac{3 }{4 \tau^2} \left( 1 + \beta \,\epsilon \right) {\rm d} \tau^2 \,.
\label{K3_mod_spm}
\end{equation}
In this case $a>0$ for each toric CY we have analysed, and so the metric is regular at the boundaries of $\mc{M}_r$. This implies that $\mc{M}_r$ is compact. In fact, its volume looks like:
\begin{equation}
\Delta \phi = \frac{\sqrt{3}}{2} \sqrt{1 +\beta \,\epsilon}\ln\left( \frac{f_\1(\vo_\0,\tau_\0)}{a\,\tau_\0}\right) \,,
\label{DphiK3}
\end{equation} 
where the homogeneous function $f_\1$ can take two different forms:
\begin{enumerate}
\item[($i$)] $f_\1(\vo_\0,\tau_\0) = b \, \vo_\0^{2/3} + \mc{O}(\epsilon)$ 

\item[($ii$)] $f_\1(\vo_\0,\tau_\0) = b \, \vo_\0/\sqrt{\tau_\0}$
\end{enumerate}
where $b>0$. Clearly (\ref{DphiK3}) satisfies the general distance bound (\ref{Delta_bound}). 

\item \textbf{SSC-like vacua} with $n_\ddP=1$ and $n_\K3f=0$:

Starting from the overall volume (\ref{vol:ssc}), if the flat direction of the reduced moduli space is parameterised by $\tau=\gamma_1\tau_s +\gamma_2\tau_*$, 
the line element of $\mc{M}_r$ takes the same form as in (\ref{dsSSC}). However this case is different since $a>0$ for each toric CY we have analysed. Thus the singularity of the metric at $\tau=0$ is outside the reduced moduli space, implying that $\mc{M}_r$ is compact. The field range $\Delta\phi$ here takes the same form as in (\ref{DphiSSC}) but with an additional subleading negative contribution coming from the lower boundary of $\mc{M}_r$.
\end{itemize}

For the remaining \textbf{structureless} LVS threefolds, it is in general not possible to derive explicit formulae for the metric on $\mc{M}_r$ and the associated field range since the hypersurface eq.~\eqref{eq:hypersurface} is equivalent to an irreducible polynomial of degree 6 in the divisor volumes, and so it cannot be solved explicitly for one of these volumes. However, we can still claim that ${\rm Vol}(\mc{M}_r)$ has to be finite via the following argument. For each structureless case, we checked that ($i$) the hypersurface (\ref{eq:hypersurface}) intersects precisely two faces of the K\"ahler cone for all values $\vo_\0>0$ and $\tau_\0 > 0$, and ($ii$) the full metric on the interior of two faces of the cone is non-singular.\footnote{With interior of the faces we mean the faces without the edges defined by the intersections of two faces.} This implies that also the induced metric on the two boundary points of $\mc{M}_r$ has to be non-singular, and so $\mc{M}_r$ has to be compact since it can be parametrised by a finite interval. 

\section{Implications}
\label{Concl}

In this section we discuss phenomenological and conceptual implications of our result.

\subsection{Phenomenological implications}

Our result has several implications for inflationary cosmology and moduli stabilisation:
\begin{itemize}
\item \textbf{Detectable tensor modes}

The number of e-foldings of inflation between horizon exit $\phi_*$ and the end of inflation $\phi_{\rm end}$ is given by:
\begin{equation}
N_e = \int_{\phi_{\rm end}}^{\phi_*} \sqrt{\frac{8}{r(\phi)}} \,{\rm d}\phi\,,
\label{Ne}
\end{equation}
where $r$ is the tensor-to-scalar ratio. Present CMB data constrain $r(\phi_*) \lesssim 0.1$ \cite{Ade:2015xua} while future cosmological observations should be able to probe values of $r(\phi_*)$ of order $0.05$ \cite{Cabass:2015jwe}. We stress that a detection of primordial gravity waves requires, strictly speaking, a large tensor-to-scalar ratio just at horizon exit. Hence, in models where $r(\phi)$ varies significantly during inflation, $r$ could be of order $0.01$ at horizon exit and then quickly decrease during the final $40$-$50$ e-foldings. As can be easily seen from (\ref{Ne}), if $r$ is very small, $N_e$ becomes quickly very large even for a small, i.e.\ a sub-Planckian, field range. Despite this observation, let us mention that so far no sub-Planckian model with detectable $r$ has been found in the context of LVS inflationary models \cite{Conlon:2005jm, Cicoli:2008gp, Broy:2015zba, Burgess:2016owb, Cicoli:2016chb, Cicoli:2011ct, Cicoli:2015wja}. 

Thus, our K\"ahler cone bound (\ref{Delta_bound}) does not necessarily imply a model-independent upper bound on the prediction of the tensor modes. However it can lead to very interesting constraints on large classes of inflationary models. If, for example, $r(\phi)$ is roughly constant during the whole inflationary evolution, (\ref{Ne}) gives (restoring the dependence on the Planck mass):
\begin{equation}
\frac{\Delta\phi}{M_p} \simeq \frac{N_e}{2}\, \sqrt{\frac{r(\phi_*)}{2}}\qquad\text{with}\qquad \Delta\phi\equiv \phi_* - \phi_{\rm end}\,.
\label{rconst}
\end{equation}
If we now use our geometrical constraint (\ref{Delta_bound}) on $\Delta\phi$, we can obtain a theoretical upper bound on the observed tensor-to-scalar ratio. 
As shown in Fig.~\ref{histogs}, the LVS vacua which allow for the largest inflaton range are K3 fibred CY threefolds with $g_s=0.3$. In order to be very conservative, we therefore focus on this LVS subclass and use the maximal values of $\Delta\phi$ for $\vo_\0=10^3$, $10^4$ and $10^5$ which are in the right ballpark to match the observed amplitude of the density perturbations \cite{Cicoli:2008gp, Broy:2015zba, Cicoli:2016chb, Cicoli:2011ct}. The results for $N_e=50$ are presented in Fig.~\ref{FigBound0} which implies $r(\phi_*) \lesssim 0.1$-$0.2$.

\begin{figure}[h!]
\begin{center}
\includegraphics[width=0.8\textwidth]{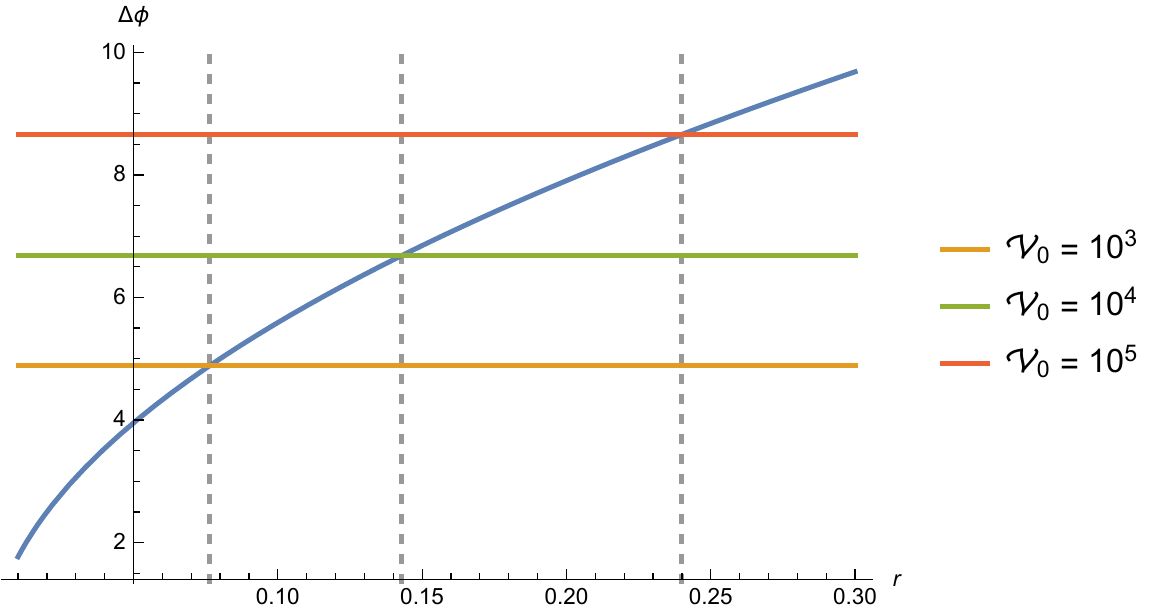}
\caption{Inflaton range $\Delta\phi$ in units of $M_p$ as a function of the tensor-to-scalar ratio $r$ for constant $r$, and corresponding upper bounds for $g_s=0.3$, $N_e=50$ and $\vo_\0=10^3$, $10^4$ and $10^5$.} 
\label{FigBound0}
\end{center}
\end{figure}

This upper bound on the value of $r$ at horizon exit is approximately of the same order of magnitude as the one derived from present CMB data which give the 
constraint $r(\phi_*) \lesssim 0.1$ \cite{Ade:2015xua}. Hence our geometrical upper bound on the inflaton range might not seem to rule out primordial gravity waves at the edge of detectability. However the results shown in Fig.~\ref{FigBound0} might very well be too naive for two main reasons: ($i$) the field range $\Delta\phi$ considered in (\ref{rconst}) does not include the post-inflationary region where the potential should develop a minimum, and so the additional requirement that also this region should be inside the CY K\"ahler cone, would necessarily make the upper bound on $r$ stronger; ($ii$) for most of the inflationary models, the approximation of constant tensor-to-scalar ratio during inflation is not very reliable. 

For example, most of the string inflation models derived in the LVS framework feature a potential which in the inflationary region can very well be approximated as \cite{Conlon:2005jm, Cicoli:2008gp, Broy:2015zba, Burgess:2016owb, Cicoli:2016chb, Cicoli:2011ct, Cicoli:2015wja}: 
\begin{equation}
V \simeq V_0\left(1-c_1\,e^{-c_2\phi}\right)\,,
\label{Vstring}
\end{equation}
which gives:
\begin{equation}
\epsilon = \frac12\left(\frac{V'}{V}\right)^2 \simeq \frac12\,c_1^2\,c_2^2\,e^{-2c_2\phi}\,.
\end{equation}
Defining the endpoint of inflation as $\epsilon(\phi_{\rm end})\simeq 1$ and the value of the tensor-to-scalar ratio at horizon exit as $r(\phi_*)=16\,\epsilon(\phi_*)$, (\ref{Ne}) can be solved explicitly to obtain (restoring again the dependence on $M_p$):
\begin{equation}
\frac{\Delta\phi}{M_p} \simeq \frac{N_e}{2}\, \sqrt{\frac{r(\phi_*)}{2}}\,\ln\left(\frac{4}{\sqrt{r(\phi_*)}}\right)\,.
\label{rStaro}
\end{equation}
Due to the extra logarithmic dependence compared with the relation (\ref{rconst}) for the case with $r$ constant, we now obtain a stronger upper bound on the predicted tensor-to-scalar ratio when we combine (\ref{rStaro}) with our geometrical constraint (\ref{Delta_bound}). Focusing again on K3 fibred LVS vacua with $g_s=0.3$ and considering the maximal values of $\Delta\phi$ for $\vo_\0=10^3$, $10^4$ and $10^5$, we get the results shown in Fig.~\ref{FigBound} for $N_e=50$. Interestingly, we now find $r(\phi_*) \lesssim 0.02$ for $\vo_\0=10^5$, $r(\phi_*) \lesssim 0.01$ for $\vo_\0=10^4$ and $r(\phi_*) \lesssim 0.005$ for $\vo_\0=10^3$. This result implies that the next generation of cosmological observations, which should be sensitive to values of $r(\phi_*)$ of order $0.05$ \cite{Cabass:2015jwe}, should not detect primordial gravity waves.

\begin{figure}[h!]
\begin{center}
\includegraphics[width=0.8\textwidth]{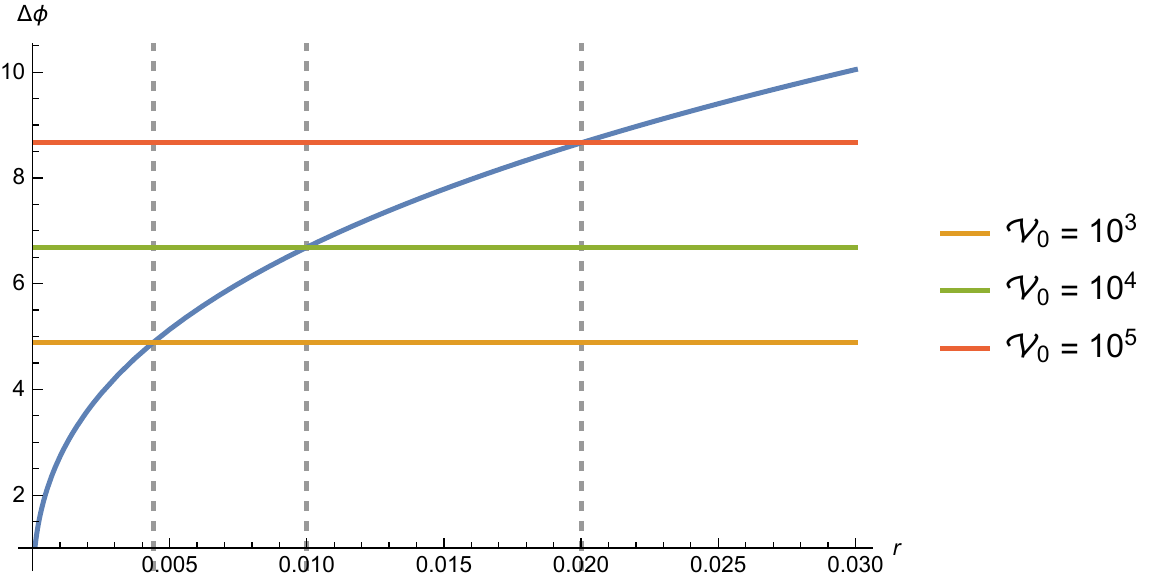}
\caption{Inflaton range $\Delta\phi$ in units of $M_p$ as a function of the tensor-to-scalar ratio $r$ for inflationary potentials of the form (\ref{Vstring}), and corresponding upper bounds for $g_s=0.3$, $N_e=50$ and $\vo_\0=10^3$, $10^4$ and $10^5$.} 
\label{FigBound}
\end{center}
\end{figure}

\item \textbf{Large and small field models}

As shown in Fig.~\ref{histolb}, LVS single-field inflationary models built from CY threefolds with $h^{1,1}=3$ split into large and small field models characterised respectively by a trans- and a sub-Planckian inflaton field range. Models with $\Delta\phi \gtrsim M_p$ feature a K3 fibration and are about $25\%$ of the total, while models with $\Delta \phi\lesssim M_p$ are based on CY threefolds without a K3 fibred structure which constitute about $75\%$ of the total number of LVS vacua with $h^{1,1}$=3 and 1 flat direction. The reason why K3 fibrations feature a much larger moduli space is that they saturate the inequality (\ref{Delta_bound}) which, in turn, follows from the form (\ref{K3_mod_spm}) of their moduli space metric. 

These purely geometric considerations, combined with the particular shape of the inflationary potential, determine the inflaton range needed to obtain enough e-foldings of inflation to solve the flatness and horizon problems. In models where the inflaton is a K3 fibre modulus, $\Delta\phi$ can be either trans-Planckian, if the inflationary potential is generated by perturbative corrections \cite{Cicoli:2008gp, Broy:2015zba, Cicoli:2016chb, Burgess:2016owb, Cicoli:2016xae, Cicoli:2017axo}, or sub-Planckian, if $V(\phi)$ is developed by non-perturbative effects \cite{Cicoli:2011ct}. On the other hand, scenarios where the inflaton is the volume of either a blow-up mode \cite{Conlon:2005jm,Cicoli:2017shd} or a Wilson divisor \cite{Blumenhagen:2012ue} are necessarily small field models where the inflationary potential can be generated only at non-perturbative level since in these cases string loop or higher derivative effects would require a trans-Planckian $\Delta\phi$ in contradiction with our analysis. 

Notice that the allowed inflaton field range depends on the stabilised values of the volume of the CY threefold $\vo_\0$ and of a diagonal dP divisor $\tau_\0$ which, in turn, depends on the string coupling $g_s$. The upper bounds presented in Fig.~\ref{histolb} are referred to the values $10^3 \leq \vo_\0 \leq 10^5$ which are in general required to reproduce the observed amplitude of the scalar modes, and $g_s=0.1$ which is a standard reference value to trust the perturbative expansion. The distribution of moduli space sizes for SSC cases do not change significantly for different values of $g_s$ since in these cases the lower bound on the inflaton range is $\tau_\0$-independent, as can be seen from the fact that $a=0$ in (\ref{eq:field_range}). Also SSC-like geometries are not very sensitive to the value of $g_s$ since their moduli space is on average smaller than the one of SSC LVS vacua. The situation is different for K3 fibred LVS vacua which, as illustrated in Fig.~\ref{histogs}, can slightly increase their allowed $\Delta\phi$ for larger values of the string coupling even if the inflaton field range can never be larger than $10$ Planck units for values of $g_s$ which are still compatible with perturbation theory.

\item \textbf{Consistency of inflationary model building}

If we restrict our analysis to large field models with an inflationary potential of the form (\ref{Vstring}) which emerge naturally in LVS compactifications, our geometrical upper bound (\ref{Delta_bound}) does not just affect the prediction of primordial tensor modes but it sets also strong constraints on the allowed underlying parameter space. In fact, a fully consistent string model should not just be able to drive enough e-foldings of inflation but should also satisfy several conditions like: ($i$) a viable global embedding into a concrete CY orientifold compactification with explicit brane set-up and fluxes compatible with tadpole cancellation; ($ii$) the presence of a chiral visible sector with a GUT or an MSSM-like gauge group whose degrees of freedom can be successfully excited at reheating after the end of inflation; ($iii$) the generation of the correct amplitude of the density perturbations by the inflaton fluctuations; ($iv$) a safe decoupling of all the modes orthogonal to the inflaton direction so that the inflationary dynamics is stable; ($v$) the minimum of the potential, and not just the inflationary plateau, should be inside the CY K\"ahler cone; ($vi$) the EFT should be fully under control throughout the whole inflationary dynamics, i.e.\ no KK mode should become light during inflation. 

Explicit chiral CY embeddings of fibre inflation models have been recently derived in \cite{Cicoli:2017axo} but not all the models are able to satisfy simultaneously all these constraints. Using the inflaton field range (\ref{DphiK3}) for a given value of $\vo_\0$, models where the form of $f_\1$ satisfy case ($ii$) seem more promising than case ($i$) for cosmological applications since they feature a larger range for the inflaton field which can be used to achieve enough e-foldings together with a large value of the tensor-to-scalar ratio of order $r\sim 0.005 - 0.01$.

\item \textbf{Curvaton-like models}

The main condition which becomes very constraining when combined with our upper bound (\ref{Delta_bound}), is the requirement to match the COBE normalisation for the amplitude of the density perturbations. In fact, if we focus again on the inflationary models described by the potential (\ref{Vstring}), the COBE normalisation condition fixes $V_0$ which is a function of the overall volume $\vo_\0$. In turn, given that the upper bound $\Delta \phi \lesssim \ln\vo$ depends on the CY volume $\vo$, the maximal size of the reduced moduli space is fixed by the observation of the amplitude of the density perturbations. This typically requires values of the internal volume of order $\vo_\0 \sim 10^3$-$10^5$, depending on the exact value of the VEV of the flux superpotential.

However multi-field models with more than one field lighter than the Hubble constant during inflation, might be characterised by non-standard mechanisms for the generation of the density perturbations. In these curvaton-like models, the inflaton fluctuations give a negligible contribution to the amplitude of the density perturbations which are instead generated by another light field \cite{Burgess:2010bz,Cicoli:2012cy}. In LVS models natural candidates for curvaton-like fields are very light axions associated with large 4-cycles. These scenarios might open up the possibility to match the COBE normalisation condition for larger values of $\vo_\0$, leading to a larger allowed field range for the inflaton field.

\item \textbf{$\alpha$-attractors}

As pointed out in \cite{Burgess:2016owb} (see in particular Fig. 3 on page 23), LVS inflationary models characterised by the scalar potential (\ref{Vstring}) represent stringy realisations of $\alpha$-attractor and Starobinsky-like models \cite{Kallosh:2013maa, Kallosh:2015zsa, Carrasco:2015pla} which seem to describe present CMB data rather well. A peculiar feature of the scalar potential (\ref{Vstring}) is the presence of an inflationary plateau which in standard $\alpha$-attractor models, derived just within the supergravity framework, is assumed to be of arbitrary length in field space. However we have shown that, when these promising inflationary models are embedded in concrete string compactifications, an additional consistency constraint should be taken into account. In fact, the K\"ahler cone of the underlying CY threefold forces the inflationary plateau to be of finite size with important implications for the prediction of crucial cosmological observables like $N_e$ or $r$.

\item \textbf{Moduli stabilisation}

Besides inflation, our results constrain also any moduli stabilisation scenario which exploits subleading perturbative or non-perturbative effects to lift the flat directions which parametrise the reduced moduli space $\mc{M}_r$. One such a scenario was discussed in \cite{Cicoli:2016chb} where these moduli were lifted in a model-independent way for any LVS vacuum by subleading $F^4$-type corrections \cite{Ciupke:2015msa}. In this case the moduli VEVs depend purely on the topological data of $X$ as well as the values of $\vo_\0$ and $\tau_\0$. Indeed, direct computation for some examples where the exact Mori cone is available shows that these vacua still lie within $\mc{M}_r$. It would also be interested to extend our results to heterotic vacua which feature an LVS-like moduli stabilisation procedure \cite{Cicoli:2013rwa}. 
\end{itemize}

\subsection{Conceptual implications}

Let us now discuss some more theoretical implications of our result:

\begin{itemize}
\item \textbf{LVS moduli space conjecture}

The main result of this paper is the proof of the compactness of the 1-dimensional reduced moduli space $\mc{M}_r$ of LVS CY threefolds with $h^{1,1}=3$. Following the intuitive picture presented in the introduction, $\mc{M}_r$ can be seen to be parameterised by a divisor which cannot collapse or grow to infinite volume since it is obstructed by the fact that the sizes of both the overall volume and a diagonal dP divisor are kept fixed. This simple understanding has led us to conjecture the validity of this result also for LVS vacua with arbitrary $h^{1,1}$ where the reduced the moduli space $\mc{M}_r$ becomes higher-dimensional. Similarly to the $h^{1,1}=3$ case, we expect CY threefolds with several K3 fibrations to feature the largest reduced moduli space also for $h^{1,1}>3$. 

\item \textbf{Comparison with weak gravity and swampland conjectures}

The formulation (\ref{Delta_bound2}) of the LVS moduli space conjecture in terms of the cut-off $\Lambda$ of the EFT treatment, allows us to compare our results with other bounds on field range excursions in moduli space which have been recently invoked by using the weak gravity and swampland conjectures \cite{Baume:2016psm,Klaewer:2016kiy,Blumenhagen:2017cxt,Palti:2017elp,Hebecker:2017lxm}. 

Our upper bound (\ref{Delta_bound2}) looks very similar to the ones derived in \cite{Baume:2016psm,Klaewer:2016kiy,Blumenhagen:2017cxt,Palti:2017elp,Hebecker:2017lxm}. However our constraint comes from purely geometric considerations associated with the size of the K\"ahler cone of the underlying CY compactification manifold, while the other bounds are basically derived by requiring a trustable EFT approach. Due to this different origin, our geometrical upper bound on the inflaton field range can be considered to be both weaker and stronger than the ones coming from the the weak gravity and swampland conjectures, depending on point of view. In fact, it can be thought to be weaker in the sense that, as the inflaton travels in field space, it would reach the point where some KK or winding modes become light before hitting the walls of the K\"ahler cone. On the contrary, our result (\ref{Delta_bound2}) can also be thought to be stronger, i.e.\ more constraining, since it sets an upper bound for $\Delta\phi$ which might hold even if the effect of heavy KK or stringy modes would be taken into account. These corrections would definitely modify both the background geometry and the definition of the correct K\"ahler coordinates, and so our upper bound (\ref{Delta_bound2}) would certainly be quantitatively modified. However we believe that our main result, i.e. the fact that the volume of the reduced moduli space is finite, would still be qualitatively correct since the exponentially large volume characteristic of LVS models should provide a powerful tool to control the effect of $\alpha'$ and KK corrections.

\item \textbf{Systematic search of new LVS vacua} 

In order to demonstrate the compactness of the LVS reduced moduli space in the 1-dimensional case, we had to determine the complete ensemble of LVS vacua with 3 K\"ahler moduli. Hence we performed a systematic search through the existing list of CY threefolds realised as hypersurfaces embedded in toric varieties \cite{Kreuzer:2000xy}. This analysis led to the discovery of several new LVS vacua compared with similar searches performed in the past \cite{Gray:2012jy,Altman:2017vzk}. However, these searches were looking for the desired structure in the intersection polynomial, and so they can be considered as complementary to our approach which is instead based on scans for divisor topologies. As the authors of \cite{Gray:2012jy, Altman:2017vzk} did not claim generality, it is not surprising that our results differ from theirs. In particular, we find 25 new LVS geometries for $h^{1,1}=3$ and 561 new LVS geometries for $h^{1,1}=4$ compared to \cite{Altman:2017vzk}. For $h^{1,1}=2$ our results precisely match with those of \cite{Altman:2017vzk}, most likely due to the simplicity of this subclass of LVS vacua. Moreover, our results are also compatible with the search for K3 fibred LVS geometries performed in \cite{Cicoli:2011it}. 

Notice also that Tab.~\ref{tab:h113} shows the presence of 43 CY geometries which have all the required properties to realise fibre inflation \cite{Cicoli:2008gp} since they feature a K3 fibred structure together with a diagonal dP divisor. This is in agreement the scanning results of \cite{Cicoli:2016xae} where a non-chiral global embedding of fibre inflation models has been presented.\footnote{Unlike the present case, the earlier scan in \cite{Cicoli:2016xae} used 526 examples corresponding to different triangulations, some of which may however correspond to the same CY geometry.}

Let us finally point out that the scanning results presented in Tab.~\ref{tab:gen_res} show that LVS vacua are highly generic in this corner of the type IIB landscape since a very large percentage of toric CY hypersurface threefolds allow for the presence of an LVS minimum, namely 56.4$\%$ for $h^{1,1}=2$, 43.4$\%$ for $h^{1,1}=3$ and 37.5$\%$ for $h^{1,1}=4$.
\end{itemize}

\section{Conclusions}
\label{Concl2}

In this paper we investigated the space of flat directions of IIB Calabi-Yau orientifold models after partial moduli stabilization in an LVS vacuum.\footnote{Note that the stabilization in an LVS vacuum is crucial to derive the field-range bounds, for other moduli stabilization proposals such as KKLT \cite{Kachru:2003aw} our argument would not imply any field-range bound.} Our main result is captured by an LVS moduli space conjecture which states that due to the CY K\"ahler cone this moduli space is compact and that its volume respects the bound in eq.~\eqref{Delta_bound2}. This bound takes a form similar to recent results in relation to the Swampland conjecture \cite{Baume:2016psm,Klaewer:2016kiy,Blumenhagen:2017cxt,Palti:2017elp,Hebecker:2017lxm}. We supported our claim by proving it for a complete subset of CY geometries, namely the $h^{1,1}=3$-Kreuzer-Skarke list giving rise to one-dimensional moduli spaces. We demonstrated how the compactness arises as a result of the CY K\"ahler cone conditions and their determination posed the main computational challenge in our analysis. To this end it was necessary to determine the ensemble of LVS vacua. By searching for diagonal del Pezzo divisors for $h^{1,1}=3,4$ we found 586 previously unknown CY LVS-geometries. In total roughly every second CY geometry in this corner of the landscape of toric hypersurface threefolds admits at least one LVS-vacuum.

The bound in eq.~\eqref{Delta_bound2} gives rise to a new strong constraint on inflationary model building in LVS, constraining any model where the inflaton is described by a K\"ahler modulus \cite{Conlon:2005jm, Cicoli:2008gp, Broy:2015zba, Burgess:2016owb, Cicoli:2016chb, Cicoli:2011ct, Cicoli:2015wja}. The distribution of the field-ranges displayed in fig.~\eqref{histolb} and fig.~\eqref{histogs} splits into two practically disjoint pieces which correspond to CYs featuring a K3-fibration and  those without such a fibration-structure. The former class induce super-Planckian field-ranges and are, therefore, more likely to lead to an observable tensor-to-scalar ratio, while the remaining geometries feature mostly sub-Planckian field-spaces.

In the future it would be desirable to explicitly prove the conjectured compactness of the reduced moduli space of LVS vacua with $h^{1,1}>3$. 

\acknowledgments

We would like to thank Roberto Valandro and Ross Altman for useful discussions. The work of PS has been supported by the ERC Advanced Grant ``String Phenomenology in the LHC Era" (SPLE) under contract ERC-2012-ADG-20120216-320421. PS is also grateful to INFN-Bologna for hospitality during the initial stage of this work.

\bibliographystyle{JHEP}
\bibliography{Kahler_cone}

\end{document}